\documentclass{ws-ijmpe}

\begin{document}
\def\beq{\begin{equation}}
\def\eeq{\end{equation}}
\def\bea{\begin{eqnarray}}
\def\eea{\end{eqnarray}}

\newcommand{\eps}{\epsilon}
\newcommand{\veps}{\varepsilon}

\newcommand{\etal}{{\it et al.},}
\newcommand{\nn}{\nonumber}

\markboth{Aurel Bulgac and Yongle Yu}{Superfluid LDA: Local Density Approximation  
for Systems with Superfluid Correlations }

%
\catchline{}{}{}{}{}
%

\title{Superfluid LDA (SLDA): Local Density Approximation  
for Systems with Superfluid Correlations}

\author{\footnotesize Aurel Bulgac and Yongle Yu }

\address{Department of Physics, University of Washington, 
 Seattle, WA 98195-1560, USA}

\maketitle

\begin{history}
\received{(received date)}
\revised{(revised date)}
\end{history}

\begin{abstract}
 We present a concise account of our development of 
the first genuine Local Density Approximation (LDA) to the 
Energy Density Functional (EDF) for fermionic systems with superfluid 
correlations, with
a particular emphasis to nuclear systems.
\end{abstract}

\section{ General remarks }

\noindent The theorem of Hohenberg and Kohn \cite{kohn} concerning the
existence of a universal  Energy Density Functional (EDF), and its subsequent
implementation as  a Local Density Approximation (LDA), lead to a new
qualitative approach to  the study of electron systems, from atoms, to
molecules, to condensed  matter systems and macromolecules
in particular and other fermionic systems in general. Even though neither
Hohenberg and Kohn  nor Kohn and Sham gave us recipes on how to construct the
EDF, various  approximation schemes with increasing level of sophistication
have been created.  Moreover, the EDF ideology  has been extended to finite
temperatures and  finite excitation energies as well. However, essentially all
of the  implementations of the LDA and EDF have been limited so far to normal
Fermi systems,  namely, systems with no pairing correlations. There were two
attempts to  extend the LDA to superconducting systems \cite{gross}, however,
the pairing  field in this approach was still a nonlocal object. One can
present the  argument that because electron superconductivity is phonon
mediated, and  since phonons have a spectrum limited by the Debye frequency,
such a  genuinely nonlocal character of the electron pairing field is
natural. In our  opinion this kind of argumentation is somewhat tenuous. The
normal  part of the EDF arises from Coulomb interaction, which in itself has an
infinite  range. Nevertheless, a coherent LDA approach to normal systems is
possible.  In recent works 
\cite{prl02,prc02,prl03,vortex_nm,vortex_a,lectures,archive,gases} we have been
able  to develop a genuinely local extention of the LDA to systems with
superfluid  correlations and apply it to a number of nuclear and atomic
systems.   Besides the fact that a genuine local approach to pairing
correlations  within an LDA \`a la Kohn and Sham is certainly possible, such a
framework  is physically meaningful. In nuclear physics for example, the so
called coherence  length, or in other words the size of the Cooper pair, is
significantly  larger than the radius of the 
NN-interaction.  The binding energy of the Cooper pair, roughly equal to the
pairing gap, is significantly smaller than the Fermi energy. Thus a zero range
approximation to the pairing interaction is definitely a meaningful thing to
pursue, once one learns how to deal with the inherent short range/ultraviolet
divergence characteristic to any local pairing field. Not the last among
various arguments that one can bring forward, is the fact that our intuition is
so much better in the case of local potentials, when compared to the case of
nonlocal potentials. This argument alone, together with the intrinsic
simplicity of a local treatment will alone warrant the quest for a suitable,
from a pure pragmatic point of view, local scheme, even if such a scheme 
would not be possible in principle.
 
\section{Formulation of the Superfluid LDA (SLDA)}

We suggest a new acronym for the extention of the LDA to superfluid systems,
namely SLDA, standing for Superfluid LDA. The starting point of the entire
formalism is naturally the assumption that a local EDF for superfluid systems
exists, namely \cite{prl02,prc02,prl03,vortex_nm,vortex_a,lectures,archive,gases}
\bea
& & {\cal{E}}({\bf r})={\cal{E}}_N[\rho({\bf r}),\tau({\bf r})] +{\cal{E}}_S[\rho({\bf r}),\nu({\bf r})], \\
& & \rho({\bf r}) = \sum_i2|v_i({\bf r})|^2,                 \\
& &    \tau({\bf r}) = \sum_i 2|\vec{\bf \nabla} v_i({\bf r})|^2,   \\
& &    \nu({\bf r}) = \sum_i v_i({\bf r})^* u_i({\bf r}),
\eea
where ${\cal{E}}_N[\rho({\bf r}),\tau({\bf r})]$ is the normal contribution  and 
${\cal{E}}_S[\rho({\bf r}),\nu({\bf r})]$ is the superfluid
counterpart. $\rho({\bf r})$ and  $\tau({\bf r})$ are the normal density and
kinetic energy  densities and $\nu({\bf r})$ is the anomalous (superfluid)
density, all  expressed through the quasiparticle wave functions $u_i({\bf
  r}),v_i({\bf r})$.  We have not shown explicitly the spin degrees of freedom
(so  far we have limited ourselves to systems with $s$-wave pairing only). The
EDF  can and does depend on a number of other local densities, which for the
sake  of the simplicity of the presentation we choose not to display as
well. We  assume that the normal part of the EDF is known and we shall not
discuss its  origin and form.

In most physical systems the magnitude of the anomalous density is relatively
small,  which reflects the fact that in most cases the pairing is in the weak
coupling  regime. One can then safely assume that, in nuclei for example, the
superfluid EDF  is only quadratic in the anomalous density and in that case
SLDA  equations acquire the following structure (shown here only for one kind
of  fermions):
\bea
& & E_{gs} = \int d^3r \{{\mathcal{E}}_N[\rho({\bf r}),\tau({\bf r})]
   +{\mathcal{E}}_S[\rho({\bf r}),\nu({\bf r})]\}, \\
& & {\mathcal{E}}_S[\rho({\bf r}),\nu({\bf r})]:= -\Delta({\bf r})\nu_c({\bf r})
   = g_{{\it eff}}({\bf r})|\nu_c({\bf r})|^2 ,\\
& & \left \{ \begin{array}{l}
 [h ({\bf r})  - \mu] u_i ({\bf r})
     + \Delta ({\bf r})  v_i ({\bf r})
    = E_i u_i ({\bf r}) , \\ 
  \Delta^* ({\bf r}) u_i ({\bf r})  -
    [ h ({\bf r}) - \mu ] v_i ({\bf r})
     = E_i v_i ({\bf r}),
\end{array}
\right . \\
& & h ({\bf r})
 = -\vec{\bf \nabla} \frac{\hbar^2}{2m({\bf r})}\cdot\vec{\bf \nabla}
 + U({\bf r}),   \\
& &     \Delta({\bf r} )
   := -g_{{\it eff}}({\bf r})\nu_c({\bf r}), \\
& & \frac{1}{ g_{{\it eff}}({\bf r})}=
\frac{1}{g[\rho({\bf r})]} 
 -\frac{m({\bf r}) k_c({\bf r})}{2\pi^2\hbar ^2}
 \left \{ 1   -\frac{k_F({\bf r})}{2 k_c({\bf r})}
\ln \frac{k_c({\bf r})+k_F({\bf r})}{k_c({\bf r})-k_F({\bf r}) }
    \right \}     \\
& & \rho_c({\bf r}) =
     \sum _{ E_i\ge 0}^{ E_c} 2|v_i({\bf r})|^2, \\
& &     \nu_c({\it r}) =
     \sum _{ E_i\ge 0}^{ E_c} v_i^*({\bf r})u_i({\bf r}),   \\
& &    E_c  +\mu =
 \frac{\hbar^2k_c^2({\bf r})}{2m({\bf r})} + U({\bf r}), \\
& & \mu =
 \frac{\hbar^2k_F^2({\bf r})}{2m({\bf r})} + U({\bf r}). 
\eea
(NB In Ref. \cite{prl03} in the equation for the renormalized coupling constant
there is a typo and the effective mass should be used as shown above.) The
SLDA equations for the quasiparticle wave functions have exactly the same structure
as the HFB/Bogoliubov-de Gennes equations, for the trivial reason that the
corresponding EDF depends on one-body densities alone. Unlike the HFB approximation,
however, SLDA does in principle lead to the exact ground state energy and
densities,  up to gradient corrections. The gradient corrections are additional
terms in EDF determined by $\vec{\bf \nabla} \rho({\bf r})$ and $\vec{\bf \nabla}
\nu({\bf r})$, which otherwise vanish identically in infinite homogeneous matter. While
the dependence of the EDF on densities can be inferred from {\it ab initio}
calculations of infinite homogeneous matter, the gradient corrections require
additional input. The uncertainties still existing in their
determination are the source of the largest errors in various EDF
approaches and the main {\it raison d'\^etre} for the letter ``A'' in the
acronym LDA. As the vast experience accumulated over the years (mainly in the
study of a large number of electron systems) amply shows, gradient corrections are never
dominant and are indeed always (relatively small) corrections.

Above $E_c$
is a cutoff energy, which should be chosen $\approx 1.5$ times the value of the
Fermi energy or larger.
In such a case any dependence of the results/observables on the value of the
cutoff  energy $E_c$
disappears. The only new element in the formalism, when compared to a formalism
with an  explicit energy 
cutoff, is the position and energy cutoff running coupling constant $g_{\it
  eff}({\bf r})$.   The main difference
with the similar running coupling constants in Quantum Field Theory (QFT) for
example, is  the fact 
that $g_{\it eff}({\bf r})$ depends
on position as well, because we are dealing with inhomogeneous systems, unlike
the  QFT case of particles interacting 
in vacuum, which is by default homogeneous. The apparent quantum mechanical
inconsistency that one can have at the same time a dependence of the running
coupling constant on both position and energy (more exactly on momentum)
cutoff is easily resolved if one remembers that as a matter of fact there is a
clear separation of scales, similar to the separation of scales in the Landau
Fermi liquids theory for example.

\section{ Application of SLDA to atomic nuclei}
 
Our knowledge of the normal part of the nuclear EDF is more or less
satisfactory  \cite{others}.
In our calculations of nuclear properties we have used the so called SLy4
interaction  \cite{sly} in order
to generate ${\cal E}_N[\rho({\bf r}),\tau({\bf r})]$ and also Fayans'
FaNDF$^0$ \cite{fayans}, both of which where in  somewhat different ways fitted
to the canonical infinite matter  results 
\cite{nm}.  
 
As far as ${\cal E}_S[\rho({\bf r}),\nu({\bf r})]$ goes,  the theoretical
knowledge is in a very unsatisfactory  overall state. Infinite matter
calculations made within the HFB/BCS framework  lead to maximum pairing gaps of
the order of 3 MeV for  $k_F\approx 1\; fm^{-1}$ and essentially vanishing
pairing gaps at nuclear saturation  densities. Various "correlation effects"
taken {\it a posteriori} into  account are essentially never in agreement with
each other, except for the  fact that the maximum value of the gap is {\em
  reduced } to about 1 MeV, a  value which a number of people think is a
reasonable one  \cite{lombardo,clark,schwenk}. In qualitative agreement with
these results  is a somewhat less known result in the nuclear community,
established more  than four decades ago. In very dilute systems the BCS value
for the pairing gap is incorrect  and the actual value, which can be estimated
quite accurately,  is smaller by a factor of about 2.2
\cite{gorkov,heiselberg}. The situation  becomes even more confusing in a way
when one takes  into account the fact that nuclei have a surface. It was shown
that surface  modes lead to a significant {\em enhancement} of the pairing gap
\cite{terasaki}, thus just to the opposite effect.

Nuclear phenomenology does not fare much better in this respect and one can
find claims that  the nuclear pairing energy has either a volume, or a surface
or a surface +  volume character \cite{jacek,dob,dens,volsurf,gogny}. Thus the
density dependence  suggested by nuclear matter calculations (weak pairing
inside and somewhat  stronger outside) does not seem to be either confirmed or
disproved in  phenomenological studies of finite nuclei. Moreover, so far we
really have no  clue on whether pairing interaction is momentum dependent
and/or energy dependent  and/or isospin dependent. Just about the only thing we
can state with some  certainty is the fact that pairing correlations are indeed
present in nuclei  and on some kind of average the paring gap is about 1 MeV or
so. And since most  of the phenomenological studies of the pairing effects are
made with quite a  number of restrictions, widely varying from one study to
another, the  more detailed information about pairing effects is often
contradictory and  thus cannot be trusted. 

In our fully self-consistent treatment of nuclei, with an exact treatment of
the continuum  spectrum as well,
we have resorted to the most simple and meaningful approach, and we have
decided to use a  single bare coupling constant for both neutrons and
protons. We have  studied the case of $T=1$ pairing only so far. Unlike
many/most of the  existing treatments available in the literature, our approach
is fully  consistent with the isospin symmetry of the nuclear forces.
Within this framework we have been able to achieve an agreement with experiment
for the  single-nucleon and two-nucleon separation energies which compares
extremely  favorably with any of the previous calculations, see 
Refs. \cite{prl03,archive}. 

We have tried to determine whether the pairing coupling constant has any
noticeable density dependence and were unable to find any. We attribute this to
the fact that a nuclear Cooper pair is unable to resolve such fine details,
since it effectively averages the pairing interaction over the entire nuclear
volume. In a recent fully self-consistent analysis of all known nuclear masses
Goriely \etal \cite{samyn} have arrived independently at the same
conclusion. This particular aspect deserves a little more discussion. Often it
is argued that the pairing strength among nucleons should be weaker/vanishing
inside and reach its maximum at the nuclear surface and beyond. Such a behavior
follows from a naive implementation of a "local density approximation" (not to
be confused with LDA for EDF) to the pairing gaps as a function of density in
infinite homogeneous nuclear matter \cite{sb}. A better alternative to the naive 
local density approximation is either the Thomas-Fermi approximation. A discussion 
of the merits and demerits of the Thomas-Fermi approximation, which indeed would
provide only an approximate solution to the SLDA equations, was performed by Grasso and 
Urban \cite{gu} in the HFB limit. 
In Ref. \cite{gu} one can find as well a comparison between the 
renormalization scheme proposed by \cite{prl02} us and its precursor \cite{bruun}.
In the weak coupling limit,
the value of the HFB/BCS pairing gap is quite accurately given by the Emery's
formula \cite{emery}.
When one introduces the low-density corrective pre-exponential factor
\cite{gorkov,heiselberg}, the corrected HFB/BCS Emery's formula for the pairing 
gap becomes (if $m^*=m$)
\bea
\Delta& =& \left ( \frac{2}{e} \right ) ^{7/3} \frac{\hbar ^2 k_F^2}{2m}
\exp \left ( -\frac{\pi}{2 \tan \delta (k_F)}\right ) \nn \\
&\approx& \left ( \frac{2}{e} \right ) ^{7/3} \frac{\hbar ^2 k_F^2}{2m}
\exp \left ( \frac{\pi}{2mk_F V(k_F,k_F)}\right ).
\eea
In the naive local density approximation, when the value of the pairing gap at
a given point inside a nucleus is given by the value of the pairing gap in
infinite homogeneous matter at the corresponding local density, a strong
density dependence can arise if the $^1S_0$ effective pairing interaction in momentum
representation $V(k_F,k_F)$ has a strong momentum dependence. A strong momentum
dependence arises if the range of the effective NN pairing interaction is
relative large, namely of the order of $1/k_F$ or larger. In the usual HFB
approximation a density dependence of this type does also lead to a large
nonlocality of the pairing field. Notice, however, that in SLDA the pairing
field is local and has no momentum dependence. In SLDA  the momentum dependence of the
pairing field can arise only if one were to introduce a dependence of
the EDF on a new anomalous density, similar to the normal kinetic energy
density $\tau ({\bf r})$, namely
\beq
\nu_\tau ({\bf r})  = \sum_i \vec{ \bf \nabla}  v_i({\bf r})^* \cdot \vec{\bf \nabla} u_i({\bf r}).
\eeq

This simple argument, based on naive local density approximation, that a
noticeable density dependence should be present in nuclei,
is not without its merits and it is indeed surprising that our and Goriely
\etal \cite{samyn} analyses  could not find any convincing trace of such a
behavior. Even if one  is willing to accept our argument that {\em a spatially
large  and weakly bound Cooper pair cannot really resolve 
such small details} and it effectively averages the strength of the pairing 
coupling constant over the entire nuclear volume,  it is not clear why such an
average is independent of the atomic number. Clearly the ratio surface/volume
is not constant over the periodic table. 
 
It is worth noticing as well, that so far, in the somewhat limited study we
have performed, we  never found a reason to introduce an isospin dependent
coupling, even though  we have definitely covered the nuclear region where such
a dependence was  advocated rather strongly by others. Moreover, to stress it
again, both proton  and neutron pairing correlations have been described with
the same coupling  constant, as indeed one would expect (up to relatively small
Coulomb and CSB  forces corrections).

\section{ Isospin structure of the superfluid contribution to the nuclear EDF}

\noindent Even though we have not been able to identify any isospin dependence
of the pairing EDF, such  a dependence could exist and it was suggested in various ways before. 
Very often however, various authors have violated the isospin symmetry of the
pairing EDF, either  invoking phenomenological arguments, or simply in order to
be able to obtain  a better description of various nuclear properties, in
particular masses  \cite{samyn,goriely,pearson}. We suspect that one can
reconcile to some  extent (not fully though) the lack of isospin symmetry in
Refs.  \cite{samyn,goriely,pearson} for example, in a rather simple manner, by
simply noticing  that so far experimentalists have been able to create more
neutron  rich nuclei than proton rich nuclei. It can be shown that 
\beq
\left \langle \frac{N-Z}{A} \right \rangle  = 0.1473, 
\eeq
where the average is computed over all measured nuclear masses with $A\geq 8$
used in Ref. \cite{goriely}.  One can then easily show that the following superfluid EDF
\bea
& & {\cal{E}}_S(\rho_n,\rho_p,\nu_n,\nu_p)\nn \\
& & =
g(\rho_n,\rho_p) \left [ |\nu_n|^2 +|\nu_p|^2 \right ] +
f(\rho_n,\rho_p) \left [ |\nu_n|^2 -|\nu_p|^2 \right ] \frac{\rho_n-\rho_p}{\rho_n+\rho_p},\\
& & g(\rho_n,\rho_p)= g(\rho_p,\rho_n) < 0,\quad f(\rho_n,\rho_p)= f(\rho_p,\rho_n) > 0, \\
& &    \frac{f(\rho_n,\rho_p)}{g(\rho_n,\rho_p)}\approx - 0.39
\eea
will reproduce the fact than on average the proton pairing is stronger than the
neutron pairing, in  agreement with the otherwise isospin violating treatment
of pairing correlations  in Refs. \cite{goriely,pearson,samyn}. The above nuclear superfluid 
EDF has two parts. The first part, proportional to $(|\nu_n|^2 +|\nu_p|^2)$, is
indeed isospin symmetric.  However, the second part, which is proportional to
$(|\nu_n|^2 -|\nu_p|^2)(\rho_n-\rho_p)$,  is only charge symmetric.\footnote{
AB is grateful to  J. Dobaczewski for helping him in clarifying this issue.}
It is highly debatable,  however, whether one can really accept a charge
symmetric only (as opposed  to an isospin symmetric one) contribution to the
nuclear superfluid EDF,  merely for the sake of improving the agreement of the
calculated masses for  example with the experimental values.

So far we could not find a suitable candidate for a nuclear superfluid EDF,
which could in principle  directly couple the proton and neutron superfluids
and which is not more  than quadratic in the anomalous densities. There have
been suggestions for  superfluid nuclear EDF in literature \cite{twofluids},
quartic in character. Recently this question has been raised again
\cite{buckley}, in order  to determine whether protons in neutron stars form a
type I or  type II superconductor, following the observations of a long period
precession  in isolated pulsars
 \cite{link}, which apparently do not support the standard picture of protons
being a type  II superconductor. We find the superfluid EDF suggested in
Ref. \cite{buckley},  however, very hard to reconcile with our knowledge of
the pairing  correlations at the corresponding densities \cite{lombardo}.

\section{Concluding Remarks} 

\noindent A relatively simple in structure and very easy to implement LDA to
EDF of fermion  systems with superfluid correlations has been developed, which
in itself  apparently represents the first genuinely local extension of the
Kohn-Sham ideology  to such systems.

This Superfluid LDA (SLDA) has been applied by us so far to study
single-nucleon and two-nucleon  separation energies of a relatively large
number of nuclei (more  than 200) with a surprisingly high accuracy. We have
used the simplest  possible ansatz for the superfluid energy density,
compatible with all basic  nuclear symmetries, and have used standard forms for
the normal  energy density part of the nuclear EDF. Namely, we have used a
superfluid EDF  characterized by a single, density independent and universal
bare coupling  constant.
And even though we did not try to obtain the best overall description of these
basic energy  nuclear properties, which are affected most significantly by the
pairing  correlations, our results proved to be of a better quality than any of
the  previous results we are aware of. Of course, there is no guarantee that by
extending  the analysis to more nuclei the accuracy of this simple approach
will survive  unscathed. But there is plenty of room for "improvements," if such would be needed.

There is no question in our minds that a more sophisticated form of the
superfluid EDF is going  to be needed in order to improve even further the
quality of the agreement  between theory and experiment. In particular, it is
very likely  that an explicit isospin dependence of the pairing couplings will eventually emerge. 

There are a number of questions our results left so far unanswered, for various
reasons.  In particular, it is still unclear whether a contribution to the
nuclear  superfluid EDF exists, which couples directly the proton and neutron superfluids. 

The microscopic calculations available in literature, concerning the dependence
of the  pairing gaps on density leave so far a lot to desire, as no consensus
seem to have  emerged to what is the true value of the pairing gap for example in neutron matter 
\cite{lombardo,clark,schwenk}. Moreover, one can expect that the values and the
density dependence  of the pairing gaps could be entirely different in
symmetric nuclear matter,  if one were to naively extrapolate the arguments of
Heiselberg \etal  \cite{heiselberg} to finite densities, and we should expect
an enhancement of  the pairing correlations when compared to simple BCS/HFB
calculations, thus  an effect opposite to that established so far in pure
neutron matter. The  dependence of the pairing gap on the isospin composition
of the matter is  still unknown microscopically. Things are however somewhat
worse, as an  additional density dependence should appear in finite
nuclei. Since nuclei have  clear and well defined surface collective modes and
as it appears  that their contribution to the pairing gaps are about 50\% or so
\cite{terasaki},  this leads us to be believe that gradient density corrections
to the  superfluid EDF could be rather large. 

Some of these aspects of the nuclear superfluid  EDF become particularly acute
in neutron stars,  as at higher than nuclear saturation densities the pairing
in $p$- and $f$-waves  becomes important. For example, it is a matter of
current debate whether at  such densities protons are a superconductor of type
I or of type II  \cite{buckley,link}, and the answer to this question can
change a lot of  the neutron stars physics.

The SLDA formalism developed by us has been applied to other systems as well,
the vortex state in  low density supefluid neutron matter \cite{vortex_nm}, the
vortex in a  superfluid dilute atomic Fermi gas \cite{vortex_a}, overall
properties of  superfluid correlations in such systems \cite{gases},
2-dimensional quantum  dots \cite{2dim}, in many cases leading to new
qualitative findings.  The range of phenomena to which SLDA has been applied so
far, with  pairing gaps spanning almost twenty orders of magnitude, is a
measure of its  flexibility and relevance.   

It is worth mentioning that there are at least two cases in which the SLDA
approach can be  implemented already in a fully controlled manner. Namely, in
the low  density regime both the normal EDF and the pairing gap are known to a
high degree  of accuracy \cite{gorkov,heiselberg}. Unlike the HF approach,
which would be  valid in the leading order in this case, the HFB/BCS fails in
this limit,  however, SLDA works. The HF approximation is accurate for
electrons in high  density regime as well. As a matter of fact we are not aware
of any physical  system in any regime where the unadulterated HFB/BCS
approximation will have  a satisfactory accuracy. Another extremely interesting
and  universal regime is that of a dilute system, but with an infinite
scattering length  \cite{george}, a regime which is nowadays routinely achieved
in dilute  atomic gases. Quite accurate calculations of such homogeneous
systems became  available recently 
\cite{carlson1,carlson2} and subsequently the SLDA approach was implemented for
inhomogeneous  systems, specifically to describe the vortex state,  see Ref. \cite{vortex_a}.

It is fair to conclude that our knowledge of the pairing properties of both
finite nuclei  and infinite neutron and nuclear matter are in a very
unsatisfactory state.  The heavily phenomenological trend, which dominated the
study of  pairing nuclear properties over the last forty years left us with no
answers to  most of the questions discussed here. It is our hope that the
existence of  a theoretically consistent SLDA framework should be of
significant help  in settling some of these aspects. 

A fundamental aspect of the nuclear pairing problem concerns the calculations
of nuclear  masses. The emphasis in nuclear physics so far was to generate,
following  one strategy or another, see
Refs. \cite{samyn,goriely,pearson,neural}, the  best possible nuclear mass formula. 
Our attitude should change, from trying to describe these masses with the best
possible  accuracy, to trying to understand why, when using a meaningful
formalism, which  incorporates our best established physical input, we still
fail at some  level or another. 
We should realize that by now we have at our disposal close to 2,500 measured
masses and  we should treat this as an object in its entirety, from which we
may learn  new physics perhaps, by using a framework consistent with what we
have  learned so far about nuclear interactions. 
 
\section*{Acknowledgments}

There is a long list of people from whom the first author (AB) has benefitted 
in various (sometimes indirect) ways over the last few years, when work on this
project  took place, namely:  
P.F. Bedaque, G.F. Bertsch, R.A. Broglia, J. Carlson, J. Dobaczewski,
B. Friman, E.K.U. Gross,  N. Hayashi,  P.-H. Heenen,
E.M. Henley, Yu. A. Litvinov, W. Nazarewicz, Yu.N. Novikov, P. Magierski,
J. Meng, G. M\"unzenberg, V.R. Pandharipande, J.M. Pearson, P. Pizocherro,
P.-G. Reinhard, C. Scheidenberger, B. Spivak, S. Stringari, A. Svidzinsky,  
N. Takigawa, E. Vigezzi, H.A. Weidenm\"uller and some of the referees of our 
papers.  This work would have been almost impossible without the financial
support of DOE  under the contract DE-FG03-97ER41014.

\end{document}